# Deflating Quantum Mysteries via the Relational Blockworld

## Abstract

We present a spacetime setting for non-relativistic quantum mechanics that deflates "quantum mysteries" and relates non-relativistic quantum mechanics to special relativity. This is achieved by assuming spacetime symmetries are fundamental in a blockworld setting, i.e., by interpreting spacetime relations as fundamental to relata. To justify this *Relational Blockworld* (RBW), we adopt a result due to G. Kaiser whereby the relativity of simultaneity, stemming from the Poincaré algebra of special relativity, is responsible for the canonical commutation relations of non-relativistic quantum mechanics. And, we incorporate a result due to A. Bohr, B. Mottelson and O. Ulfbeck whereby the density matrix for a given experimental situation is obtained from its spacetime symmetry group. We provide an example to illustrate the explanatory nature of RBW and conclude by explaining how RBW deflates "quantum mysteries."

# 1. Introduction

In this article we provide a spacetime setting for non-relativistic quantum mechanics (QM) that relates QM and special relativity (SR) while deflating the so-called "quantum mysteries" as explained pedagogically by Mermin[1], Greenberger *et al*[2], Kwiat & Hardy[3], Aravind[4], and Laloë[5] among others. The interpretation of QM presented herein assumes spatio-temporal relations (as described by the spacetime symmetry group) are fundamental in a blockworld setting as justified by the formalism of QM itself. That is, we understand quantum facts to be facts about the spatiotemporal relations of a given physical system, not facts about the behavior of particles or the interactions of measurement devices with wave-functions, etc. According to the view in which relations are fundamental in a blockworld – which we call the *Relational Blockworld* (*RBW*) – reality is fundamentally relational and non-dynamical. It is our view that quantum phenomena such as non-locality and non-separability are conceptually problematic only when we attempt to formulate a dynamical explanation for something that is irreducibly relational and non-dynamical. On our view, *all* phenomena are "non-separable" via the spatiotemporal holism of RBW. Thus, since RBW ties quantum non-separability and non-locality, long thought to pose serious potential problems for the relativity of simultaneity, to a spacetime where blockworld is true, RBW deflates "quantum mysteries" in a straightforward fashion.

We begin our explanation of RBW in section 2 with an introduction to the blockworld (BW) as implied by the relativity of simultaneity (RoS) per SR. Specifically, we show how RoS implies that the future, past and present are equally 'real'. *Prima facie* RoS would not seem to bear on *non-relativistic* quantum mechanics, but in section 3 we show how QM's canonical commutation relations follow from RoS per Kaiser[6] and Bohr & Ulfbeck[7] (see also Anandan[8]). Having justified a BW setting for QM, we show in section 4 that spatiotemporal symmetries may be viewed as fundamental in this setting per the work of Bohr, Ulfbeck and Mottleson[9]. Specifically, they have shown that the density matrix of QM may be derived from the spatiotemporal symmetries of an experimental configuration alone. [The mathematical details of sections 2, 3 and 4 are not necessary for a conceptual understanding of RBW, so they are relegated to appendices.] That QM may be understood to reside in a blockworld wherein spacetime symmetries are fundamental establishes RBW and implies that spatiotemporal relationalism is a more fundamental means of explanation than dynamism.

Section 5 provides an RBW interpretation of a simple QM experiment to contrast the dynamical interpretation of this experiment with that of spatiotemporal relationalism. [The reader does not have to understand all the mathematical details of section 5 to appreciate the RBW perspective conveyed therein.] Drawing on the lessons of the entire article, section 6 then explains how RBW deflates the "quantum mysteries." Simply put, there is no mystery associated with non-relativistic quantum mechanics since non-separability, non-locality and the solution to the measurement problem are all straightforward consequences of the spacetime symmetry structure in a blockworld.

**2. Relativity of Simultaneity and the Blockworld**

The first step in establishing RBW is to understand and appreciate the BW as implied by RoS. In the next section, we show that it is precisely this "nonabsolute nature of simultaneity[10]" which survives the c → ∞ limit of the Poincaré group, and is *responsible for* the canonical commutation relations of *non-relativistic* quantum mechanics. To illustrate RoS and BW, consider the following example adopted with minor modifications from DeWitt and Mermin[11]. [Calculations are in Appendix A.]

The adaptation we make to the DeWitt/Mermin version of this story is to consider only local observations. We do this to emphasize that the BW implication of RoS is not an optical illusion resulting from the finite speed of light. In order to keep our observations local we add three new characters – Bob, Alice and Kim – to Joe and Sara of the DeWitt/Mermin version. The boys, Bob and Joe, are at rest with respect to each other and the girls are at rest with respect to each other. Joe and Bob see the girls moving in the positive x direction at 0.6c (Figure 1). The girls, therefore, see the boys moving in the negative x direction at 0.6c (Figure 4). Who is *actually* moving?

The answer to this question is central to the BW perspective. According to the first postulate of SR, there is no way to discern absolute, uniform motion so *either perspective is equally valid*. The girls are correct in saying it is the boys who are moving, and the boys are correct in saying it is the girls who are moving. This is equivalent to saying there is no preferred, inertial frame of reference in the spacetime of SR. Now we compare some observations and their consequences.

Joe is located at x = 0 (lower case coordinates are the boys') and Bob is at x = 1000km. Joe says Sara's clock read T = 0 as she passed him (upper case coordinates are the girls'). Joe's clock likewise read t = 0 at that event. Bob said Kim's clock read T = -0.0025s when she passed him. Bob's clock read t = 0 at that event (Figure 1). The girls agree with these mutual clock readings, so what's the problem?

The boys synchronized their clocks so the events Joe/Sara (event 1) and Bob/Kim (event 2) are simultaneous, having both occurred at t = 0. Clearly, according to the boys, the girls' clocks are not synchronized. Unfortunately, the girls also synchronized their clocks so, according to the girls, events 1 and 2 are *not* simultaneous. Who's right?

Neither of their frames is preferred, so the girls *and* the boys are correct! Whether or not space-like separated events (Figure 2) are simultaneous is relative to the frame of reference. So, what is the consequence of this "relativity of simultaneity?"

RoS renders the view known as "presentism" highly suspect. Presentism is the belief that everyone shares a unique, 'real' present state while their common past states no longer exist and their common future states are yet to exist. Per presentism, everyone could agree with a statement such as, "Sam is surfing in California while Jonathan is shoveling snow in New York." If Sam is 25 years old, the 24 year-old version of Sam no longer exists, i.e., is no longer 'real', and the 26 year-old version will not exist, i.e., will not be 'real', for another year. There is a sense that we share in a 'real' present moment with everything else in the universe and this attribute of 'realness' moves along all worldlines synchronously (Figure 3). In fact, what one means by "the universe" is vague unless everyone agrees on a spatial surface of simultaneity in spacetime. But, as we continue with our example, it should become clear that RoS is contrary to this commonsense notion of presentism. Let us continue.

Since Kim's clock read T = -0.0025s at event 2, the girls say Bob passed Kim 0.0025s *before* Joe got to Sara at T = 0 (Figure 4). The event simultaneous with event 1 is Bob passing Alice at T = 0, i.e., event 3 (Figure 5). All agree that Alice's clock read T = 0 when Bob was there, but they also agree that Bob's clock read t = 0.002s when he was at Alice's position. Here's how the girls tell the story.

Sara is at X = 0, Alice is at X = 800km and Kim is at X = 1250km. The boys are not 1000km apart, as they claim, but rather only 800km apart (this difference in spatial separation is called *length contraction*). The girls know this since Bob was at Alice's position (X = 800km, T = 0) when Joe was at Sara's position (X = 0, T = 0). As a consequence, Bob passed Kim (event 2) *before* Joe got to Sara (event 1). In fact, it took Bob 0.0025s to get from X = 1250km to X = 800km moving at 0.6c, that's why Alice's clock read -0.0025s when Bob was there. So, who shares the attribute of 'realness', i.e., where is the spatial frame of 'realness' which defines "the universe?"

Well, unless we can touch things which are not 'real', Joe and Sara are 'co-real' at event 1, and Bob and Kim are 'co-real' at event 2 (Figure 6). But, Joe and Bob are 'co-real' at events 1 & 2 (t = 0 for both) so we see that all four characters involved in events 1 & 2 are 'co-real'. This means Sara shares 'realness' at T = 0 with Kim at T = -0.0025s, *and* Sara shares 'realness' at T = 0 with Kim at T = 0. Thus, Kim at T = 0 shares 'realness' with Kim at T = -0.0025s so *RoS implies the past is as 'real' as the present*. Now let's look at the boys' perspective.

Joe is at x = 0 and Bob is at x = 1000km. Sara and Kim passed the boys (events 1 & 2) at t = 0, so Joe and Bob do not believe Sara and Kim are 1250km apart, but only 1000km apart (again, length contraction). Alice passed Bob 0.002s later, so she must be 0.6c x 0.002s = 360km behind Kim (not 450km as the girls claim). At event 3, Alice shares 'realness' with Bob while she shares 'realness' with Sara (both T = 0, Figure 6). However, at T = 0 Sara also shares 'realness' with Joe (both at event 1). Thus, Bob at t = 0 shares 'realness' with Bob at t = 0.002s so *RoS implies the future is as 'real' as the present*.

In BW all observers' histories are treated democratically, unless we 'add something' to pick out a preferred frame. No frame is physically distinguished from any other. No set of measurement records, derived in any frame of reference, is more veridical than any other. With this radical democracy of histories comes a radical democracy of spatiotemporal events. The essence of BW is that all observers' futures, pasts and presents are equally 'real'. This is the most parsimonious and least metaphysically odious hypothesis that explains the spatiotemporal events measured by all the observers in the case above, and for any case based on the geometric facts of SR.

If all spatiotemporal events are 'there', then nothing comes into being. There are no objective physical "processes over time." Events only appear to us to unfold over time, but fundamentally, nothing unfolds: all events just "are," timelessly so. This suggests that all fundamentally dynamical explanations, including all causal explanations that invoke 'interactions' or any dynamical processes, are for pragmatic purposes only. Nothing interacts in BW, nothing 'comes to be and passes away'. There is no foliation-independent reality to temporal distinctions and dynamical processes.

Of course, such dynamical notions are *logically compatible* with BW, for we can always model 'becoming' or the 'unfolding' of events in a universe that lacks such a reality at its fundamental level, just as I can represent a moving car *with a series of still photographs*. The 'moving car' is now 'nothing but' a 'series' of photos, which themselves have no 'motion' anywhere in them. To the extent that there is motion, it is now a *relation between fundamentally motionless objects*. Only when I *relate* the photos in a certain way does the 'motion' of a single "trans-temporal object[12]" arise amidst the photos. But it is nowhere 'in' the photos. Thus, I can always construct a model of change in a changeless context (without engendering a contradiction). This is simply to assert that the *appearance* of change is logically compatible with the *reality* of changelessness. But we see immediately that the change, just like the motion, is not fundamental to that which is representing it. Just like there is no motion 'in' the pictures, even though when I flash them quickly 'a car' appears 'to move', *there is no change or becoming in BW even though there appears to be*.

BW is fundamentally non-dynamical, but admits dynamical representations. That is, we represent the world as unfolding, relative to a perspective within the BW, so as to make predictions. A prediction is simply the generation of a *possible* trajectory through space and time based on our ignorance of the complete BW. Thus, all *fundamental* explanations in a BW are essentially *non-dynamical*. Of course, one can always construct dynamical accounts of phenomena to suit one's already dynamical intuitions; but dynamical accounts of some quantum phenomena will be greatly vexed if we assume that dynamical processes are fundamental. However the very same phenomena, when given a relational explanation in BW, will seem rather natural.

From the BW perspective, one may explain phenomena via correlations between events which are distributed through the past, present and future (spatiotemporal relationalism) *without regard for telling causal stories*. Thus, one needn't find a common cause to explain space-like separated, correlated events since *the spatiotemporal relation itself is* the explanation[13]. As applied to QM, this implies one needn't construct worldlines between source(s) and detectors(s) so one needn't invoke particles or waves as causal mechanisms for detector clicks. Since, as we will show in the next section, RoS is foundational to QM, it follows that interpretations of QM have at their disposal this non-dynamical potential afforded by the RBW perspective.

**3. Blockworld and Non-relativistic Quantum Mechanics**

Notice in the preceding example that if Joe 'jumps into' Sara's frame of reference at event 1 then moves spatially to Kim (Figure 6), he does not get to event 2 because he will be at the T = 0 version of Kim while the Kim of event 2 is at T = -0.0025s. If, rather, Joe moves spatially to Bob *then* 'jumps into' the girls' frame of reference, he is at the T = -0.0025s version of Kim. That is to say, Lorentz boosts (changes to moving frames of reference according to the Poincaré group of SR) do not commute with spatial translations since different results obtain when the order of these two operations is reversed. Specifically, this difference is a temporal displacement which is key to generating a BW. This is distinct from Newtonian mechanics whereby time and simultaneity are absolute per Galilean invariance. If spacetime was Galilean invariant, the boys and girls in our example would all agree as to which events were simultaneous and could subscribe to presentism. In such a spacetime, it wouldn't matter if you Galilean boosted then spatially translated, or spatially translated then Galilean boosted. *Prima facie*, one might suspect that *non-relativistic* quantum mechanics would be in accord with Galilean spacetime, but this is not the case.

Kaiser[14] has shown that the non-commutivity of Lorentz boosts with spatial translations is *responsible for* the non-commutivity of the QM position operator with the QM momentum operator. He writes[15],

> "For had we begun with Newtonian spacetime, we would have the Galilean group instead of [the restricted Poincaré group]. Since Galilean boosts commute with spatial translations (time being absolute), the brackets between the corresponding generators vanish, hence no canonical commutation relations (CCR)! In the [c → ∞ limit of the Poincaré algebra], *the CCR are a remnant of relativistic invariance where, due to the nonabsolute nature of simultaneity, spatial translations do not commute with pure Lorentz transformations*." [Italics his. A technical overview of his argument is presented in Appendix B.]

Bohr & Ulfbeck[16] also realized that the "Galilean transformation in the weakly relativistic regime" is needed to construct a position operator for QM, and this transformation "includes the departure from simultaneity, which is part of relativistic invariance." Specifically, they note that the commutator between a "weakly relativistic" boost and a spatial translation results in "a time displacement," which is crucial to RoS. Thus they write[17],

> "For ourselves, an important point that had for long been an obstacle, was the realization that the position of a particle, which is a basic element of nonrelativistic quantum mechanics, requires the link between space and time of relativistic invariance."

So, *the essence of QM – its canonical commutation relations – is a "remnant" of RoS*. This fundamental result makes it natural to invoke BW for interpreting QM and suggests a deep unity between SR and *non-relativistic* quantum mechanics[18]. This result leaves but one step in motivating RBW, i.e., to establish spatiotemporal symmetries as fundamental in QM.

**4. Spatiotemporal Relations are Fundamental in QM**

Bohr, Mottelson and Ulfbeck[19] have shown how QM can be grounded in the symmetries of spacetime. [An overview of their calculations is in Appendix C.] Specifically, the density matrix can be obtained via the spatiotemporal relationships of the experimental configuration (via the spacetime symmetry group) rendering concepts such as the Hamiltonian, mass and Planck's constant ancillary. In such a view the detector clicks are not caused by impinging particles; in fact they're not caused by any*thing*, and QM simply provides the distributions of uncaused clicks. Bohr & Ulfbeck[20] call this the *Theory of Genuine Fortuitousness* and write[21],

> "It would appear, however, that the role of symmetry in relation to quantal physics has, so to speak, been turned upside down, and it is the purpose of the present article to show that quantal physics itself emerges when the coordinate transformations (the elements of spacetime symmetry) are recognized as the basic variables."

In other words, and this is what RBW supplies to a view such as Genuine Fortuitousness, one may view spatiotemporal relations as fundamental to relata in QM so one doesn't need to employ particles to explain click distributions. Thus, we suggest that QM is best understood in the context of RBW, i.e., *spatiotemporal relations are fundamental in a BW setting*. Bohr, Mottelson and Ulfbeck write[22], "Indeed, atoms and particles as things are phantasms (things imagined)." This is where explanation 'bottoms out'.

At the fundamental level we don't specify 'fundamental objects' with dynamical laws to explain phenomena, but rather we explain phenomena via spatiotemporal relations. The deeper facts are the spacetime relations, not the things that can possibly stand in those relations. Thus QM probabilities are not probabilities for the occurrence of intrinsic events, but rather QM probabilities are an expression of the spacetime symmetries inherent in the experiment itself. In the radically relational and non-dynamical perspective of RBW, we must remember that there are just 'static' relations from which we might construct stories of trans-temporal objects participating in dynamical interactions. As an example of the RBW perspective, we consider a simple QM experiment.

## 5. Example

We re-evaluate the Mach-Zehnder interferometer (MZI) example provided by Bohr & Ulfbeck[23] (BU) while attempting to minimize the potential for dynamical inference. While we are dealing with a single frame of reference (rest frame of the experimental apparatus), the spacetime symmetries responsible for a description of the experiment will provide the distribution of outcomes. Thus, spatio-temporal relationships become the fundamental elements of description, replacing the dynamical notion of particles/waves emitted by the source, moving through the equipment and impinging on the detector(s) under the governance of dynamical laws (like the Schrödinger equation).

The starting point in the construction of our MZI is a source and detectors. Sources and detectors are mutually defined as the detector won't register an event, i.e., 'click', without the source and the source doesn't click at all. Our particular configuration involves a source-detector combination for which direction is important. In the source-detector configuration of Figure 7, only detector 1 (D1) will be active, i.e., register clicks. If the source is rotated 90° CCW, only detector 2 (D2) will register clicks, etc. The dynamical explanation for the clicks is of course that the source is emitting particles of some sort, e.g., electrons, photons, neutrons, to which the detector is sensitive. The RBW explanation is simply that the source stands in a particular spatiotemporal relation to the detectors.

To characterize that aspect of the source-detector combo that makes it 'click', we introduce a real number k with dimension of inverse length. The absolute value of k is denoted $k_o$ and called the *wave number*, $2\pi/\lambda$. In the dynamical perspective, $k_o$ corresponds to a property of the emitted particles while per RBW $k_o$ is simply an element in a mathematical expression of a spacetime symmetry (see next paragraph). [Of course, RBW does not require we abandon the language of dynamism where it is *convenient* for technical discussion.]

There are two symmetry variables that will be needed to describe the distribution of clicks in the detectors of our interferometer, i.e., 1-dim spatial translations (x → x + a) and reflections in that spatial dimension (x → –x). The irreducible representations for 1D translations and reflections in the eigenbasis of translations are

$$\mathbf{T}(a) = \begin{pmatrix} e^{-ik_o a} & 0 \\ 0 & e^{ik_o a} \end{pmatrix} \qquad \mathbf{S}(a) = \begin{pmatrix} 0 & e^{-2ik_o a} \\ e^{2ik_o a} & 0 \end{pmatrix}$$

respectively (BU's eqn. 8). [These, rather than particles or waves, are the fundamental constituents in an RBW description of the experiment.] The orientation of a source and its active detectors will be described by the eigenkets of **T(a)**, i.e., $|+\rangle$ for orientation along the positive x axis and $|-\rangle$ for orientation along the negative x axis. Figure 8 depicts a source-detector configuration described by $|+\rangle$ and Figure 9 a source-detector configuration described by $|-\rangle$.

The eigenkets of **S(a)** in the **T(a)** basis are

$$|S(a) = 1\rangle = \frac{1}{\sqrt{2}}\left(e^{-ik_o a}|+\rangle + e^{ik_o a}|-\rangle\right)$$

for eigenvalue +1, and

$$|S(a) = -1\rangle = \frac{1}{\sqrt{2}}\left(e^{-ik_o a}|+\rangle - e^{ik_o a}|-\rangle\right)$$

for eigenvalue –1 (BU's eqn. 21), as can be seen by explicit computation.

A source-detector combo that is described by |S(0) = 1> can be created by introducing a partially reflecting mirror called a beam splitter (BS, with 1/8 wavelength phase plate) as shown in Figure 10.

The partially reflecting mirror BS1 resides in the x = 0 plane. The source and detector 1 (D1) define a line whose angle with respect to BS1 is reflected to create a line from BS1 to detector 2 (D2). Both detectors are active and register the same number of clicks per unit time on average. The eigenket of **S(0)** for this configuration is

$$|S(0) = 1\rangle = \frac{1}{\sqrt{2}}(|+\rangle + |-\rangle)$$

Accordingly, the probability of registering a click in D1 is $|\langle+|S(0) = 1\rangle|^2 = \frac{1}{2}$ and that in D2 is $|\langle-|S(0) = 1\rangle|^2 = \frac{1}{2}$, where $\langle+|$ is chosen for clicks in D1, because D1 is oriented along the positive x axis with respect to the source. The effect of introducing BS1 was to change the configuration $|\psi_a\rangle = |+\rangle$ to $|\psi_b\rangle = |S(0) = 1\rangle$. We can define a unitary operator **Q(a$_o$)** which instantiates this change mathematically, i.e., (BU's eqn. 43)

$$Q(a_o) \equiv \frac{1}{\sqrt{2}}(I - iS(a_o))$$

where $a_o \equiv \pi/(4k_o)$. [Since $k_o = 2\pi/\lambda$, we see that $a_o = \lambda/8$.] Specifically,

$$Q(a_o) = \frac{1}{\sqrt{2}}\left[\begin{pmatrix} 1 & 0 \\ 0 & 1 \end{pmatrix} - i\begin{pmatrix} 0 & -i \\ i & 0 \end{pmatrix}\right] = \frac{1}{\sqrt{2}}\begin{pmatrix} 1 & -1 \\ 1 & 1 \end{pmatrix}$$

and

$$|\psi_b\rangle = Q(a_o)|+\rangle = \frac{1}{\sqrt{2}}\begin{pmatrix} 1 & -1 \\ 1 & 1 \end{pmatrix}\begin{pmatrix} 1 \\ 0 \end{pmatrix} = \frac{1}{\sqrt{2}}\begin{pmatrix} 1 \\ 1 \end{pmatrix}.$$

The next step in building our interferometer is to introduce a pair of mirrors M between BS1 and the detectors as in Figure 11.

Again, D1 and D2 register the same number of clicks per unit time on average and the experimental configuration is described by |S(0) = 1>. The difference is that now lines from the source to each detector via BS1 and M suggest the probability of clicks in D1 is given by $|\langle-|S(0) = 1\rangle|^2 = \frac{1}{2}$ while that in D2 is given by $|\langle+|S(0) = 1\rangle|^2 = \frac{1}{2}$, i.e., the mirrors M introduced a reflection about x = 0. Mathematically,

$$|\psi_c\rangle = S(0)|\psi_b\rangle = S(0)|S(0) = 1\rangle = \begin{pmatrix} 0 & 1 \\ 1 & 0 \end{pmatrix} \begin{pmatrix} \frac{1}{\sqrt{2}} \\ \frac{1}{\sqrt{2}} \end{pmatrix} = \frac{1}{\sqrt{2}} \begin{pmatrix} 1 \\ 1 \end{pmatrix} = +1|S(0) = 1\rangle$$

If we now introduce another partially reflecting mirror BS2 in the x = 0 plane, we find our configuration is that of Figure 12 where only one detector is active.

The mathematical description for $|\psi_c\rangle \rightarrow |\psi_d\rangle$ describing Figure 12 is given by $Q^\dagger(a_o)|\psi_c\rangle$. That is

$$Q^+(a_o)|\psi_c\rangle = Q^+(a_o)|S(0) = 1\rangle = \frac{1}{\sqrt{2}} \begin{pmatrix} 1 & 1 \\ -1 & 1 \end{pmatrix} \begin{pmatrix} \frac{1}{\sqrt{2}} \\ \frac{1}{\sqrt{2}} \end{pmatrix} = \begin{pmatrix} 1 \\ 0 \end{pmatrix} = |+\rangle = |\psi_d\rangle$$

This follows trivially since $Q(a_o)|+\rangle = |S(0) = 1\rangle$ (BU's eqn. 44) and $Q(a_o)$ is unitary, i.e., $Q(a_o)Q^\dagger(a_o) = I$ (BU's eqn. 43).

We can reactivate D2 if we introduce a phase plate P as shown in Figure 13. The distribution of clicks in D1 and D2 is now a function of the phase $k_o a$ introduced by the phase plate, i.e., $\cos^2(k_o a)$ in D1 and $\sin^2(k_o a)$ in D2. Thus, the introduction of the phase plate P is equivalent to introducing a translation $T(a)$ of $|\psi_c\rangle$ to obtain $|\psi_e\rangle$, then obtaining $|\psi_f\rangle = Q^\dagger(a_o)|\psi_e\rangle$. We have

$$|\psi_e\rangle = T(a)|\psi_c\rangle = \begin{pmatrix} e^{-ik_o a} & 0 \\ 0 & e^{ik_o a} \end{pmatrix} \begin{pmatrix} \frac{1}{\sqrt{2}} \\ \frac{1}{\sqrt{2}} \end{pmatrix} = \frac{1}{\sqrt{2}} \left( e^{-ik_o a}|+\rangle + e^{ik_o a}|-\rangle \right)$$

which is an eigenket of $S(a)$ per

$$S(a)|\psi_e\rangle = \begin{pmatrix} 0 & e^{-2ik_o a} \\ e^{2ik_o a} & 0 \end{pmatrix} \begin{pmatrix} \frac{e^{-ik_o a}}{\sqrt{2}} \\ \frac{e^{ik_o a}}{\sqrt{2}} \end{pmatrix} = +1 \begin{pmatrix} \frac{e^{-ik_o a}}{\sqrt{2}} \\ \frac{e^{ik_o a}}{\sqrt{2}} \end{pmatrix} = +1|\psi_e\rangle$$

Now to obtain $|\psi_f\rangle$ for Figure 13 from $|\psi_e\rangle$ we have $|\psi_f\rangle = Q^\dagger(a_o)|\psi_e\rangle$ or

$$Q^+(a_o)|\psi_e\rangle = \frac{1}{\sqrt{2}} \begin{pmatrix} 1 & 1 \\ -1 & 1 \end{pmatrix} \begin{pmatrix} \frac{e^{-ik_o a}}{\sqrt{2}} \\ \frac{e^{ik_o a}}{\sqrt{2}} \end{pmatrix} = \frac{1}{2} \begin{pmatrix} e^{ik_o a} + e^{-ik_o a} \\ e^{ik_o a} - e^{-ik_o a} \end{pmatrix} = \cos(k_o a)|+\rangle + i\sin(k_o a)|-\rangle = |\psi_f\rangle$$

(BU's eqn. 46). Note that this was obtained using only the eigenbasis of the symmetry operator **T(a)**. Likewise, <**T(a)**> suffices to specify the "state of affairs" via the density operator, since

$$\langle\psi_f|T(a)|\psi_f\rangle = \begin{pmatrix}\cos(k_o a) & -i\sin(k_o a)\end{pmatrix}\begin{pmatrix}e^{-ik_o a} & 0 \\ 0 & e^{ik_o a}\end{pmatrix}\begin{pmatrix}\cos(k_o a) \\ i\sin(k_o a)\end{pmatrix} = e^{-ik_o a}\cos^2(k_o a) + e^{ik_o a}\sin^2(k_o a)$$

tells us if we perform this experiment varying the phase we would be able to ascertain the probabilities $p(T_1)$ and $p(T_2)$, where $T_i$ are the eigenvalues of **T(a)**, by simply associating clicks in D1 and D2 (Figure 13) with $T_1$ and $T_2$, respectively. [Remember, we're not measuring any 'thing' here, so we're not concerned by imaginary eigenvalues. Our analysis differs in this respect from the dynamical analysis of BU and they write[24], "The quantum is registered by a detector D, by which the reflection symmetry S appears with the value s = +1 (signal) or s = -1 (absence of signal)."] Thus, our experimental outcomes would establish a density operator in the eigenbasis of **T(a)** given by

$$\rho = \cos^2(k_o a)|+\rangle\langle+| + \sin^2(k_o a)|-\rangle\langle-|\ .$$

Or, conversely, we could infer $|\psi_f\rangle$ up to phase via

$$\langle T(a)\rangle_\psi = T_1|\psi_1|^2 + T_2|\psi_2|^2$$

where $\psi_i$ are the components of $|\psi_f\rangle$ in the eigenbasis of **T(a)**. Either way, Bohr, Mottelson and Ulfbeck's idea that experimental outcomes of the symmetry operators determine the "state of affairs" is illustrated in this case and, therefore, the Hamiltonian **H** is merely a 'tag along'.

To explore the *dynamical* aspect of this experiment, we lapse into the realm of trans-temporal objects, re-introducing the concept of mass and its scaling factor, $\hbar$. Bohr & Ulfbeck write[25]

> "The need for the concept of mass in classical physics is thus seen to arise from the low resolution, which hides the wave number and frequency, of dimensions $L^{-1}$ and $T^{-1}$ that lie behind the conserved dynamical quantities denoted by momentum and energy. These quantities were, therefore, given dimensions $MLT^{-1}$ and $ML^2T^{-2}$, in terms of a dimension M apparently not reducible to spacetime. With the discovery of the underlying quantal structure, the two scales could be identified as having the universal ratio $\hbar$. The choice of units with $\hbar = 1$ thus eliminates the need for a dimension of mass."

That is, $\left[\dfrac{momentum}{\hbar}\right] = \dfrac{1}{length} = [wavenumber]$ and $\left[\dfrac{energy}{\hbar}\right] = \dfrac{1}{time} = [frequency]$.

The Hamiltonian for our example is

$$H = \begin{pmatrix} \dfrac{\hbar^2 k_o^2}{2m} & 0 \\ 0 & \dfrac{\hbar^2 k_o^2}{2m} \end{pmatrix}$$

when a particle of mass m is moving through the MZI, or

$$H = \begin{pmatrix} \hbar k_o c & 0 \\ 0 & \hbar k_o c \end{pmatrix}$$

when a photon is moving through the MZI. Accordingly, we see that **H** commutes with both **T(a)** and **S(a)**, meaning the energy of the quantum is invariant with respect to x → x + a (spatial translations) and x → − x (spatial reflections). Thus, we may understand that **H** is a multiple of **I** per Schur's lemma[26]. Trivially, we compute <**H**> from the density operator

$$\langle H \rangle = Tr(\rho H) = Tr\begin{pmatrix} \cos^2(k_o a) & 0 \\ 0 & \sin^2(k_o a) \end{pmatrix}\begin{pmatrix} H_1 & 0 \\ 0 & H_2 \end{pmatrix} = H_1 \cos^2(k_o a) + H_2 \sin^2(k_o a)$$

But, since $H_1 = H_2 = E$, i.e., **H** is a multiple of **I**, then we have (again, trivially)

$$\langle H \rangle = E\cos^2(k_o a) + E\sin^2(k_o a) = E$$

as would be the case for any observable that 'tags along', i.e., commutes with both **T(a)** and **S(a)**.

Thus we have illustrated, albeit with a simple example, how the density operator can be constructed via the irreducible representations of symmetry group elements alone. Consequently, we see that the Hamiltonian plays but a secondary role in the description of this experiment. Having justified and illustrated the RBW perspective, we're ready to show how it demystifies non-relativistic quantum mechanics.

## 6. Conclusion: Quantum Mysteries Deflated

RBW is a blockworld in which spatiotemporal relations are fundamental. A blockworld is a spacetime in which the future, past and present are equally 'real'. Thus, presentism is false in a BW and there is no uniquely "evolving universe" or "unfolding now." Every event that will happen or has happened just 'is' in a BW. That is, the wave-function *qua* state-space representation of QM is a calculational device. The relational spacetime symmetries of an experimental arrangement (that give rise to quantum statistics) are the deeper ontological story of QM. On our view the measurement problem is merely an artifact of the state space formalism.

Quantum non-locality and non-separability are likewise demystified in a straightforward fashion since RBW assumes spatiotemporal relations are fundamental in a BW. Correlations between space-like separated events that violate Bell's inequalities are of no concern as long as spatiotemporal relations in the experimental apparatus warrant the correlations. There is no need to satisfy the common cause principle, since non-local correlations are not about "particles" impinging on measuring devices. Rather, the non-local correlations derive from the spatiotemporal relations in the construct of the experiment. There are no influences, causal mechanisms, etc., because non-locality is a relational property that is precisely described by the spacetime symmetries of any given experimental arrangement. Nothing happens in a relational blockworld, so there is nothing for such inherently dynamical processes and entities to do.

The conceptual trouble with quantum non-locality, quantum non-separability or entanglement is a consequence of our apparent dynamical perspective 'within' a relational blockworld. The trouble is with us, not the world, so to speak. In trying to explain the spatiotemporal distribution of detector clicks as caused by or as determined by impinging particles (carrying with them their own properties), standard accounts of QM assume a Galilean background spacetime in which quantum states evolve. However, we have seen that the Galilean temporal transformation, in which time is absolute, is replaced in Kaiser's spacetime of QM by a time transformation which introduces a temporal displacement. The kinematical consequence of this temporal displacement, i.e., RoS, is "essential" for a geometrical interpretation of QM while the dynamical consequence, i.e., a phase factor, is inconsequential. Thus the non-Galilean nature of the QM spacetime is hidden in standard (dynamical) accounts of QM creating such mysteries as posed by entanglement and non-locality. Dynamical beings in a relational blockworld might only be able to explicate this non-locality in terms of "backwards causation," or something like Bohm/deBroglie's pilot-wave, but no such causal mechanisms are necessary as the BW, with spacetime symmetries, is all that is necessary to accommodate QM.

In conclusion, we realize 1) the canonical commutation relations of QM follow from RoS, which implies BW, and 2) the spatiotemporal distribution of quantum events follows from the spatiotemporal relationships of the experimental configuration. Thus, we are not forced to paint any kind of dynamical picture about quantum non-locality and non-separability. Rather, we understand quantum facts to be facts about the spatiotemporal relations of a given physical system, not facts about the behavior of particles, or the interactions of measurement devices with wave-functions, etc. Per an RBW explanation, the spacetime symmetries plus the boundary conditions (initial and final) determine the click distributions.

In some sense, Bohr was right: there is no "quantum world." But, what Bohr did not realize, and where Einstein failed as well, is that buried in the fabric of special relativity itself is the quantum – now understood to be a consequence of the non-commutivity of Lorentz boosts with spatial translations. And thus, like Einstein's SR, this is a principle as opposed to a constructive account of QM, i.e., *RBW is not a kind of instrumentalism* – on our view spacetime symmetries are not merely calculational devices, they are the foundational "elements" of reality. Just as Minkowski spacetime of SR subsumes the Galilean spacetime of Newtonian mechanics, it also subsumes Kaiser's spacetime of QM. Dynamical talk is simply a crude way of trying to describe global, static, spatiotemporal dependency relations between various regions of spacetime in a given experimental situation. According to RBW, reality is fundamentally relational and non-dynamical, so quantum phenomena such as non-locality and non-separability are conceptually problematic only when we attempt to formulate a dynamical explanation for something that is irreducibly relational and non-dynamical. In fact, *all* phenomena are "non-separable" via the spatiotemporal holism of RBW.

**Appendix A: Lorentz Transformations for Section 2**

The speed of the boys relative to the girls is 0.6c, so

$$\sqrt{1-\frac{v^2}{c^2}} = \sqrt{1-\frac{0.36c^2}{c^2}} = 0.8$$

and

$$\gamma = \frac{1}{0.8} = 1.25.$$

With $T = t = 0$ at $X = x = 0$, the girls' coordinates at the event labeled by the boys as $t = 0$, $x = 1000$km are given by the following Lorentz transformations

$$T = \gamma\left(t - \frac{vx}{c^2}\right) = 1.25\left(0 - \frac{0.6c * 1000}{c^2}\right) = -0.0025s$$

$$X = \gamma(x - vt) = 1.25(1000 - 0.6c * 0) = 1250km$$

where c = 300,000 km/s. And, the girls' coordinates at the boys' event $t = 0.002$, $x = 1000$km are

$$T = \gamma\left(t - \frac{vx}{c^2}\right) = 1.25\left(0.002 - \frac{0.6c * 1000}{c^2}\right) = 0$$

$$X = \gamma(x - vt) = 1.25(1000 - 0.6c * 0.002) = 800km$$

**Appendix B: Canonical Commutation Relations via the Relativity of Simultaneity**

[This appendix is a collection of select excerpts from Kaiser's and Bohr & Ulfbeck's extensive works.]

Take the limit $c \rightarrow \infty$ in the Lie algebra of the Poincaré group where the non-zero brackets are

$[J_i, J_n] = iJ_k$         (1,2,3 cyclic)
$[T_o, K_n] = iT_n$
$[K_i, K_n] = -(i/c^2)J_k$    (1,2,3 cyclic)
$[J_i, K_n] = iK_k$        (1,2,3 cyclic)
$[J_i, T_n] = iT_k$        (1,2,3 cyclic)
$[T_i, K_n] = -(i/c^2)\delta_{in} T_o$.

where $J_i$ are the generators of spatial rotations, $T_o$ is the generator of time translations, $T_i$ are the generators of spatial translations, and $K_i$ are the boost generators. We then obtain

$[J_i, J_n] = iJ_k$        (1,2,3 cyclic)
$[M, K_n] = 0$
$[K_i, K_n] = 0$
$[J_i, K_n] = iK_k$        (1,2,3 cyclic)
$[J_i, T_n] = iT_k$        (1,2,3 cyclic)
$[T_i, K_n] = -(i/\hbar)\delta_{in} M$

where M is obtained from the mass-squared operator in the $c \rightarrow \infty$ limit since $c^{-2} \hbar T_o = c^{-2} P_o$ and
$$c^{-2} P_o = (M^2 + c^{-2} P^2)^{1/2} = M + P^2/2Mc^2 + O(c^{-4}).$$

Thus, $c^{-2} T_o \rightarrow M/\hbar$ in the limit $c \rightarrow \infty$. ["Mass" by choice of 'scaling factor' $\hbar$.] So, letting $P_i \equiv \hbar T_i$ and $Q_n \equiv -(\hbar/m)K_n$ we have

$$[P_i, Q_n] = -\hbar^2/m [T_i, K_n] = (-\hbar^2/m)(i/\hbar)\delta_{in} mI = -i\hbar\delta_{in} I.$$

In this "weakly relativistic regime" the coordinate transformations of Appendix A now look like
$$X = x - vt$$
$$T = t - vx/c^2.$$

These differ from a Galilean transformation by the $vx/c^2$ term in T, i.e., in a Galilean transformation time is absolute, so $T = t$.

**Appendix C: Density Matrix Obtained via Symmetry Group**

Herein we present a pedagogical version of the appendix to Bohr, Mottelson and Ulfbeck[27] (BMU) wherein they show the density matrix can be derived using only the irreducible representations (irreps) of the symmetry group elements, g ε G. We begin with two theorems from Georgi[28]

> "The matrix elements of the unitary, irreducible representations of G are a complete orthonormal set for the vector space of the regular representation, or alternatively, for functions of g ε G."

This gives[29]

> "If a hermitian operator, H, commutes with all the elements, D(g), of a representation of the group G, then you can choose the eigenstates of H to transform according to irreducible representations of G. If an irreducible representation appears only once in the Hilbert space, every state in the irreducible representation is an eigenstate of H with the same eigenvalue."

What we mean by "the symmetry group" is precisely that group G with which some observable H commutes (although, these elements may be identified without actually constructing H). Thus, the mean value of our hermitian operator H can be calculated using the density matrix obtained wholly by D(g) and <D(g)> for all g ε G. Observables such as H are simply 'along for the ride' so to speak.

To show how, in general, one may obtain the density matrix using only the irreps D(g) and their averages <D(g)>, we start with eqn 1.68 of Georgi[30]

$$\sum_g \frac{n_a}{N} \left[D_a(g^{-1})\right]_{kj} \left[D_b(g)\right]_{lm} = \delta_{ab}\delta_{jl}\delta_{km}$$

where $n_a$ is the dimensionality of the irrep, $D_a$, and N is the group order. If we consider but one particular irrep, D, this reduces to the orthogonality relation (eqn 1) of BMU

$$\sum_g \frac{n}{N} \left[D(g^{-1})\right]_{kj} \left[D(g)\right]_{lm} = \delta_{jl}\delta_{km} \qquad (1)$$

where n is the dimension of the irrep. Now multiply by $[D(g')]_{jk}$ and sum over k and j to obtain

$$\sum_j \sum_k \sum_g \frac{n}{N} \left[D(g^{-1})\right]_{kj} \left[D(g)\right]_{lm} \left[D(g')\right]_{jk} = \sum_j \sum_k \delta_{jl}\delta_{km} \left[D(g')\right]_{jk} = \left[D(g')\right]_{lm}$$

The first sum on the LHS gives

$$\sum_j [D(g^{-1})]_{kj}[D(g')]_{jk} = [D(g^{-1})D(g')]_{kk}$$

The sum over k then gives the trace of D(g$^{-1}$)D(g'), so we have

$$\frac{n}{N}\sum_g [D(g)]_{lm} Tr\{D(g^{-1})D(g')\} = [D(g')]_{lm}$$

Dropping the subscripts we have eqn 2 of BMU

$$\frac{n}{N}\sum_g D(g)\, Tr\{D(g^{-1})D(g')\} = D(g'). \qquad (2)$$

If, in a particular experiment, we measure directly the click distributions associated with the various eigenvalues of a symmetry D(g), we obtain its average outcome, <D(g)>, i.e., eqn 3 of BMU

$$\langle D(g) \rangle = \sum_i \lambda_i p(\lambda_i) \qquad (3)$$

where $\lambda_i$ are the eigenvalues of D(g) and $p(\lambda_i)$ are the distribution frequencies for the observations of the various eigenvalues/outcomes. [Note: It is the experimentalist's job to assign the clicks of the detectors to eigenvalues of the symmetry operators.]

In terms of averages, BMU's eqn 2 becomes

$$\frac{n}{N}\sum_g \langle D(g)\rangle Tr\{D(g^{-1})D(g')\} = \langle D(g')\rangle \qquad (4)$$

which they number eqn 4. Since we want the density matrix to satisfy the standard relation (BMU's eqn 5)

$$Tr\{\rho D(g')\} = \langle D(g')\rangle \qquad (5)$$

it must be the case that (BMU's eqn 6)

$$\rho \equiv \frac{n}{N}\sum_g D(g^{-1})\langle D(g)\rangle \qquad (6)$$

That this density operator is hermitian follows from the fact that the symmetry operators are unitary. That is, D(g$^{-1}$) = D$^\dagger$(g) implies <D(g$^{-1}$)> = <D(g)>*, thus

$$\rho^+ = \frac{n}{N}\sum_g D^+(g^{-1})\langle D(g)\rangle^* = \frac{n}{N}\sum_g D(g)\langle D(g^{-1})\rangle = \frac{n}{N}\sum_g D(g^{-1})\langle D(g)\rangle = \rho.$$

[The second-to-last equality holds because we're summing over all g and for each g there exists g$^{-1}$.] So, the density operator of eqn 6 will be hermitian and, therefore, its eigenvalues (probabilities) are guaranteed to be real. This is not necessarily the case for D(g), since we know only that they're unitary. However, we need only associate detector clicks with the eigenvalues of D(g) and in this perspective one does not attribute an eigenvalue of D(g) to a property of some 'click-causing particle'. Therefore, whether or not the eigenvalues of any particular D(g) are real or imaginary is of no ontological or empirical concern (as was shown in the interferometer example of section 5).

Finally, to finish BMU's calculations, we note that since ρ is hermitian it can be diagonalized. BMU label its eigenvalues w(ζ) and its eigenvectors $|\zeta\rangle$ so in the eigenbasis of ρ we have (BMU's eqn 7)

$$\rho = \sum_\zeta |\zeta\rangle w(\zeta) \langle\zeta| \ . \tag{7}$$

Inserting the identity operator $I = \sum_\zeta |\zeta\rangle\langle\zeta|$ into eqn 5 yields

$$\langle D(g)\rangle = Tr\{\rho D(g)\} = Tr\left\{\sum_\zeta \rho |\zeta\rangle\langle\zeta| D(g)\right\} = \sum_m \sum_\zeta \rho_{m\zeta}[D(g)]_{\zeta m} \ .$$

In the eigenbasis of ρ we have $\rho_{m\zeta} = w(\zeta)\delta_{m\zeta}$ so we obtain the first expression of BMU's eqn 8:

$$\langle D(g)\rangle = \sum_\zeta w(\zeta)[D(g)]_{\zeta\zeta} \ . \tag{8a}$$

Inserting two identity operators expressed in the eigenbasis of D(g), i.e., $I = \sum_i |\lambda_i\rangle\langle\lambda_i|$ and $I = \sum_j |\lambda_j\rangle\langle\lambda_j|$, we obtain

$$\langle D(g)\rangle = \sum_j \sum_i \sum_\zeta w(\zeta)\langle\zeta|\lambda_i\rangle\langle\lambda_i|D(g)|\lambda_j\rangle\langle\lambda_j|\zeta\rangle = \sum_j \sum_i \sum_\zeta w(\zeta)\langle\zeta|\lambda_i\rangle\lambda_j\delta_{ij}\langle\lambda_j|\zeta\rangle$$

Simplifying gives

$$\langle D(g)\rangle = \sum_j \sum_\zeta \lambda_j w(\zeta)\langle\zeta|\lambda_j\rangle\langle\lambda_j|\zeta\rangle = \sum_j \sum_\zeta \lambda_j w(\zeta)|\langle\zeta|\lambda_j\rangle|^2$$

Comparison with eqn 3 then gives the second of BMU's expressions in eqn 8

$$p(\lambda_j) = \sum_\zeta w(\zeta)|\langle\zeta|\lambda_j\rangle|^2 \tag{8b}$$

This tells us that in general the assignment of a detector event to $\lambda_j$ does not uniquely determine its assignment to w(ζ). Thus, it is incumbent upon the experimentalist to determine specifically which click adds to which distribution function. [In the simple interferometer example this assignment was trivial, so the determination of w(ζ) was straightforward.]

To obtain the last of BMU's expressions in eqn 8 we operate on $|\zeta\rangle$ with the identity operator expressed in the eigenbasis of D(g), i.e., $I = \sum_i |\lambda_i\rangle\langle\lambda_i|$, yielding

$$|\zeta\rangle = \sum_i |\lambda_i\rangle\langle\lambda_i|\zeta\rangle \qquad (8c)$$

which is simply the expansion of $|\zeta\rangle$ in the $|\lambda_i\rangle$ basis.

**Figure Captions**

Figure 1. Picture @ t = 0. Events 1 & 2 are simultaneous according to the boys.

Figure 2. Spacetime diagram showing time-like, null and space-like separated events in a BW spacetime. Events A, B and C are time-like separated from the origin, O. Event D is space-like separated from O. Event E is null separated from O. Events A and B are space-like separated from one another so some observers will see A occur before B while others will see B occur before A. In some frame of reference A and B are simultaneous, since they are space-like separated. The same is true for events O and D.

Figure 3. Spacetime diagram of presentism. Event C occurs at time $t_{-3}$ for all observers, regardless of their relative motions. Events O, A and B occur at time $t_0$ for all observers and are therefore unambiguously simultaneous. Event D occurs at time $t_4$ for all observers. Events C and D do not exist when events O, A and B exist since C is no longer 'real' and D is not yet 'real' when O, A and B are 'real'. 'Realness' only exists at one spatial surface $t_i$ at a time. [Clearly another 'time' is needed for this last statement to make sense.]

Figure 4. Picture @ T = -0.0025s. Event 2 occurred *before* Joe gets to Sara according to the girls.

Figure 5. Picture @ T = 0. Events 1 & 3 are simultaneous according to the girls.

Figure 6. Events 1 & 2 are simultaneous for the boys (both lie along t = 0 spatial plane). Boys' spatial planes of simultaneity are horizontal (t = 0 and t = 0.002s are shown). Girls' spatial planes of simultaneity (T = 0 and T = -0.0025s are shown) are tilted relative to the boys' spatial planes, as are the girls' worldlines tilted relative to the boys' worldlines. Events 1 & 3 are simultaneous for the girls (both lie along T = 0 spatial plane).

Events 1 & 2 are 'co-real' since Joe and Bob are both at these events at t = 0. Events 1 & 3 are 'co-real' since Sara and Alice are both at these events at T = 0. Bob is at both events, so his t = 0 self is 'co-real' with his t = 0.002s self.

Figure 7. The source and detectors are directional in nature. Here, only D1 registers clicks.

Figure 8. Source and detector aligned in the positive x direction.

Figure 9. Source and detector aligned in the negative x direction.

Figure 10. A beam splitter BS1 is added to the source-detector combo causing two detectors to register clicks.

Figure 11. A pair of mirrors is introduced. The orientation of the detectors must change so that they still register clicks.

Figure 12. Another beam splitter BS2 is introduced and only one detector is active. Its orientation has changed from the previous configuration.

Figure 13. A phase plate P is introduced and detector 2 begins to register clicks.

**Figure 1**

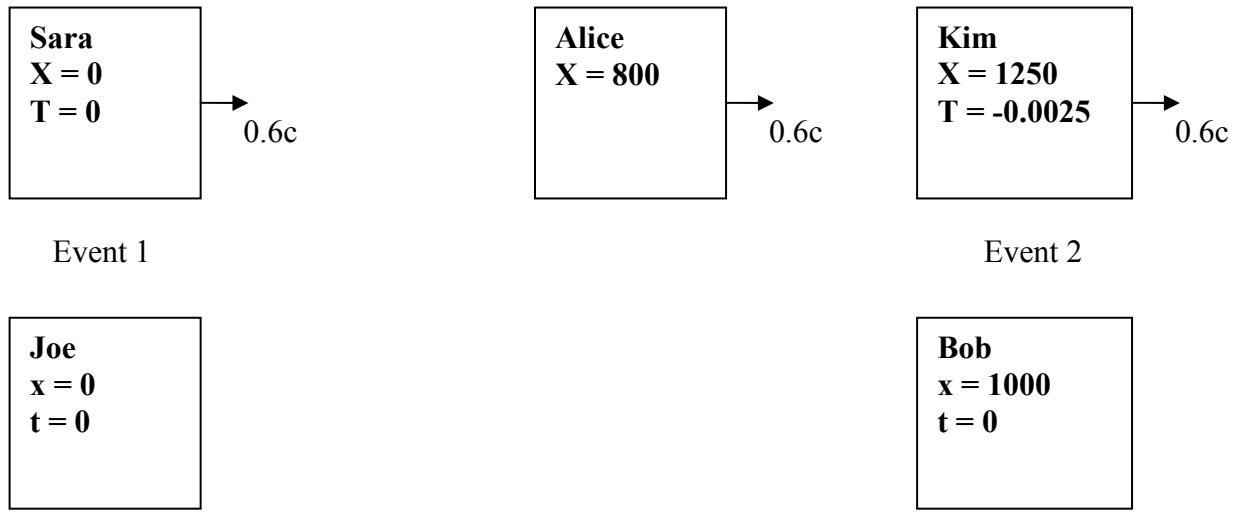

Event 1

Event 2

Sara
X = 0
T = 0
→ 0.6c

Alice
X = 800
→ 0.6c

Kim
X = 1250
T = -0.0025
→ 0.6c

Joe
x = 0
t = 0

Bob
x = 1000
t = 0

**Figure 2**

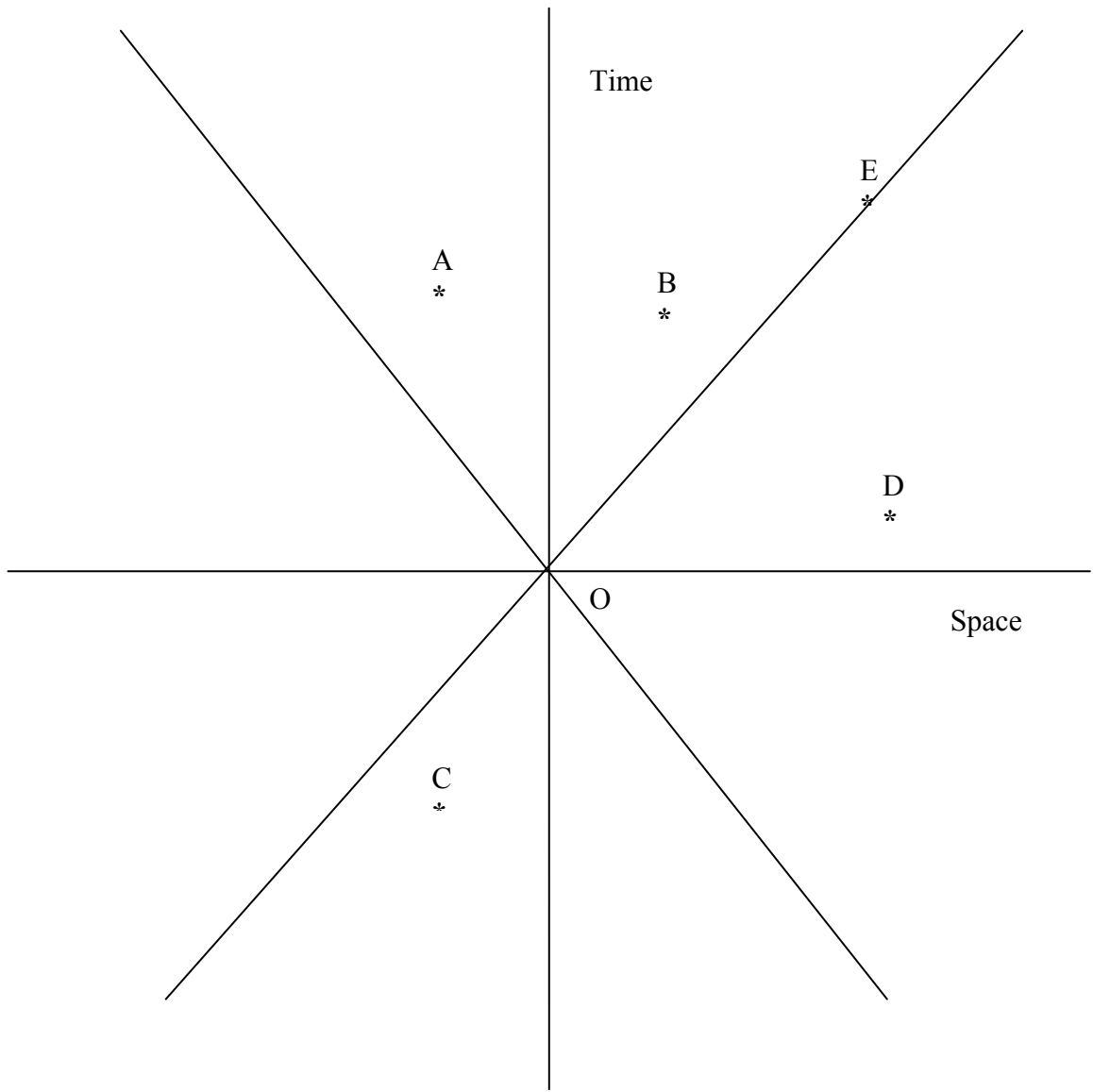

**Figure 3**

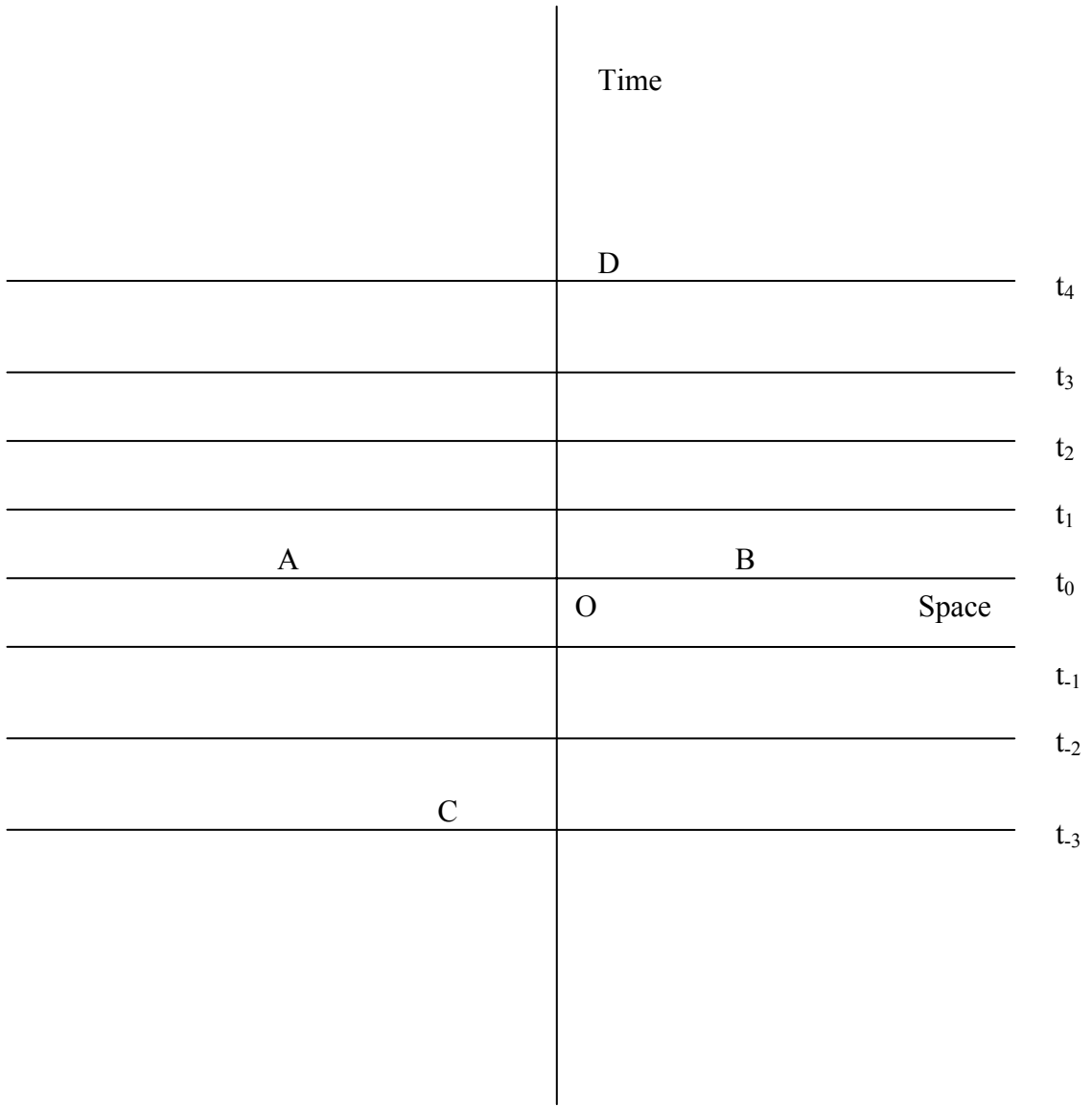

**Figure 4**

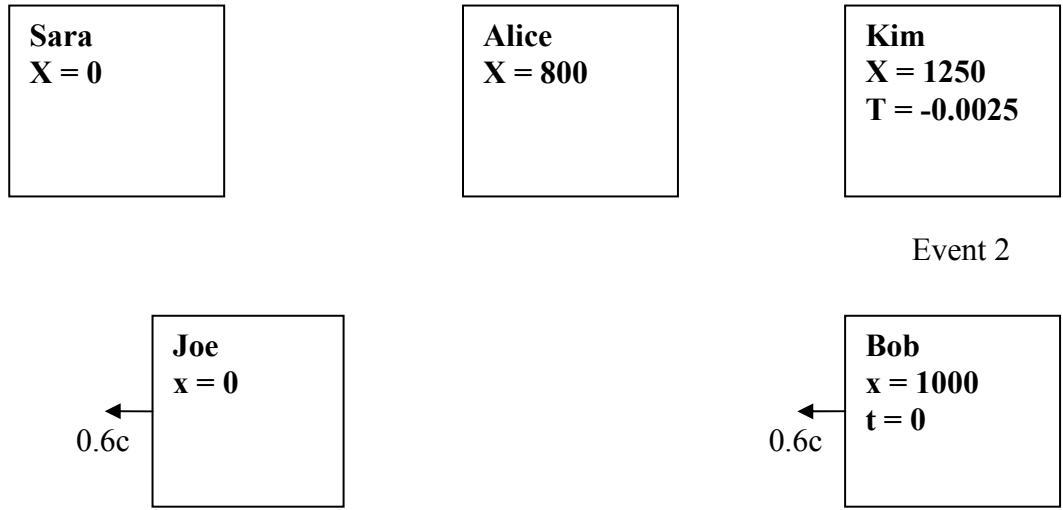

| Sara | Alice | Kim |
| X = 0 | X = 800 | X = 1250 |
| | | T = -0.0025 |

Event 2

| Joe | | Bob |
| x = 0 | | x = 1000 |
| ← 0.6c | | ← 0.6c, t = 0 |

**Figure 5**

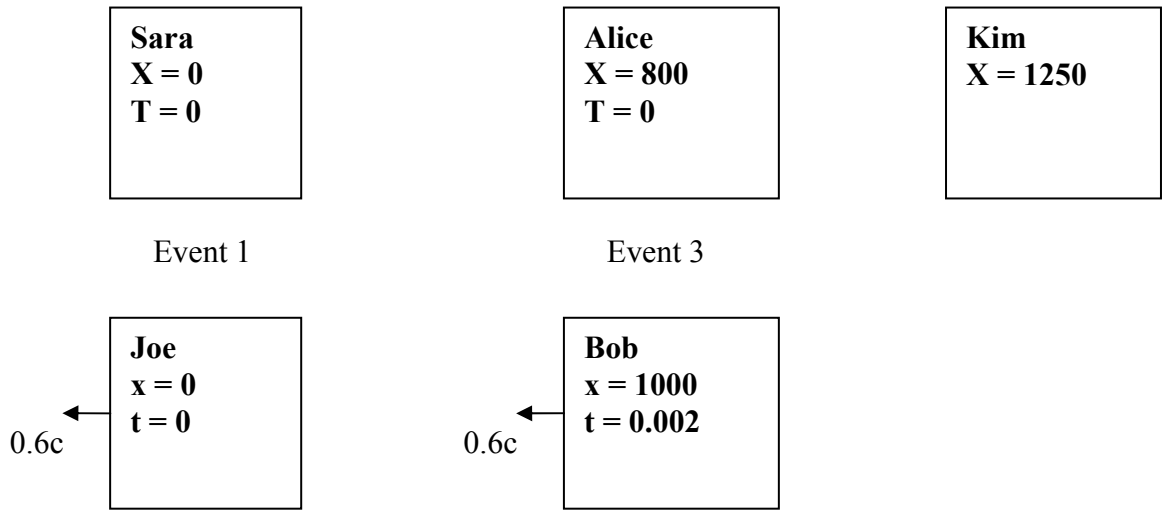

**Figure 6**

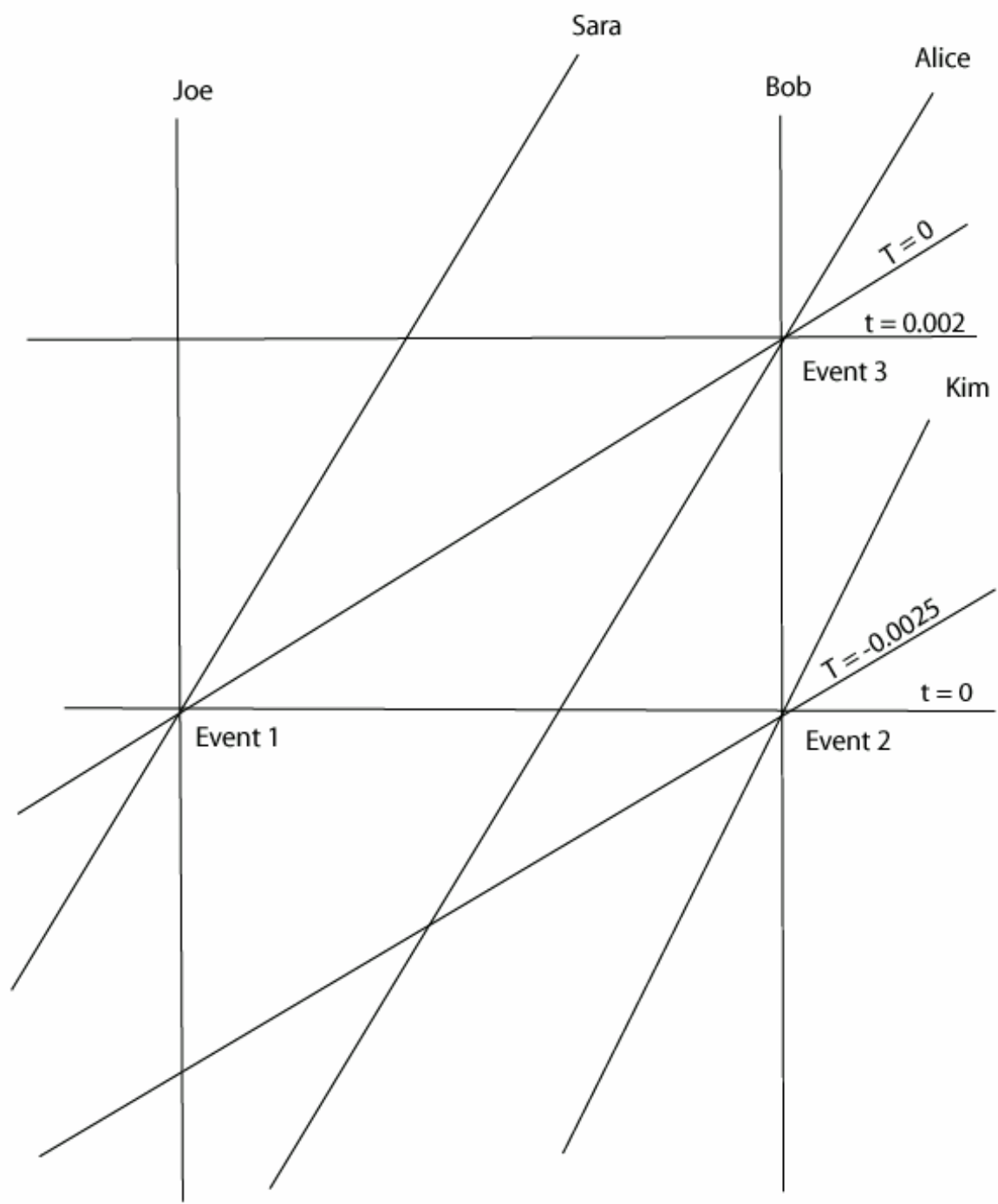

**Figure 7**

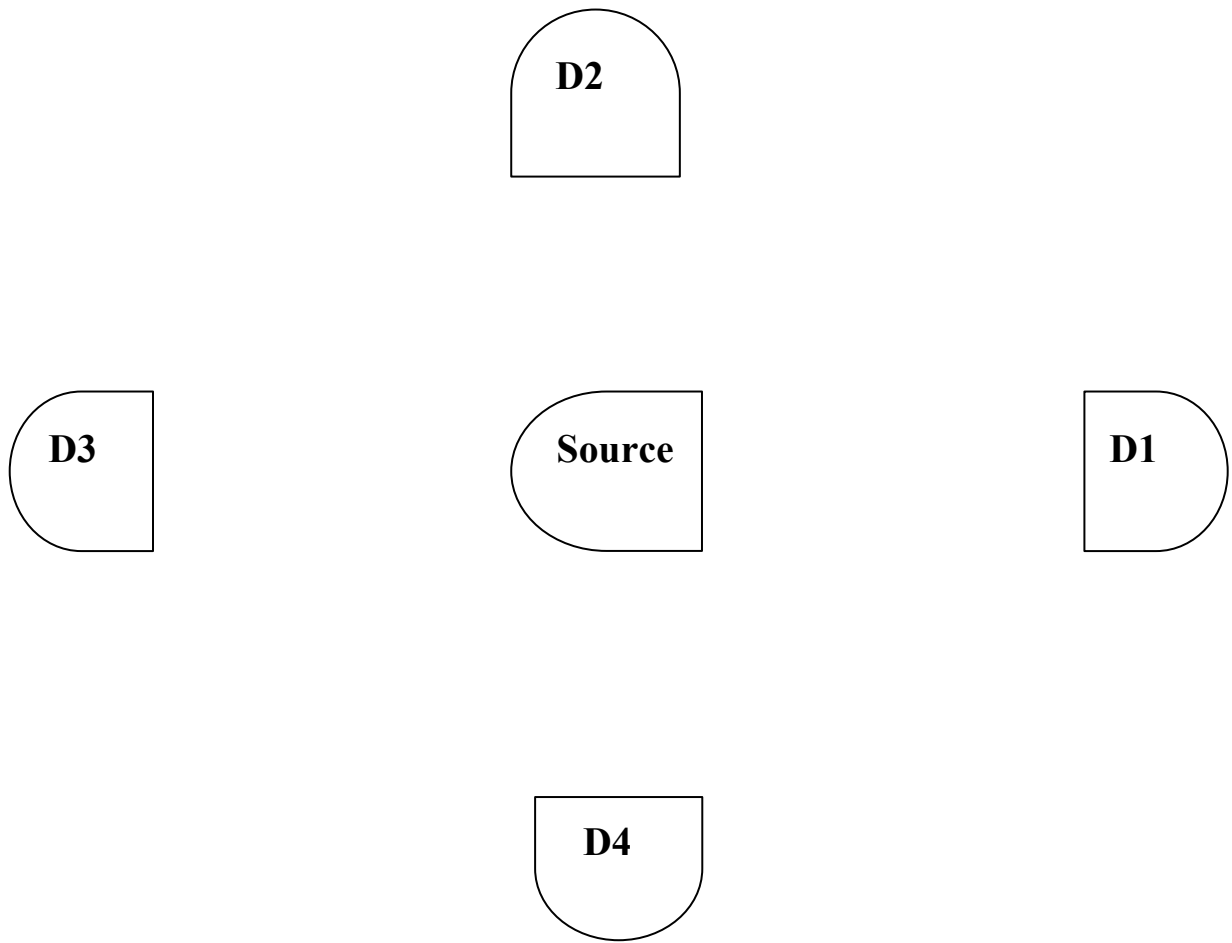

**Figure 8**

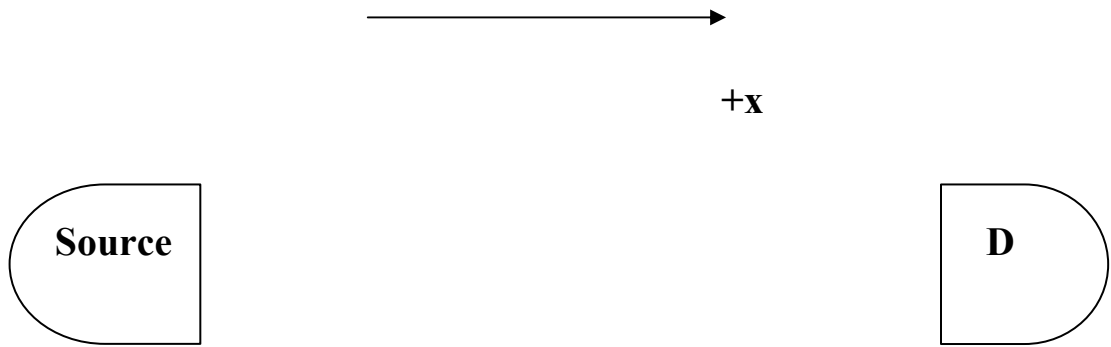

**Figure 9**

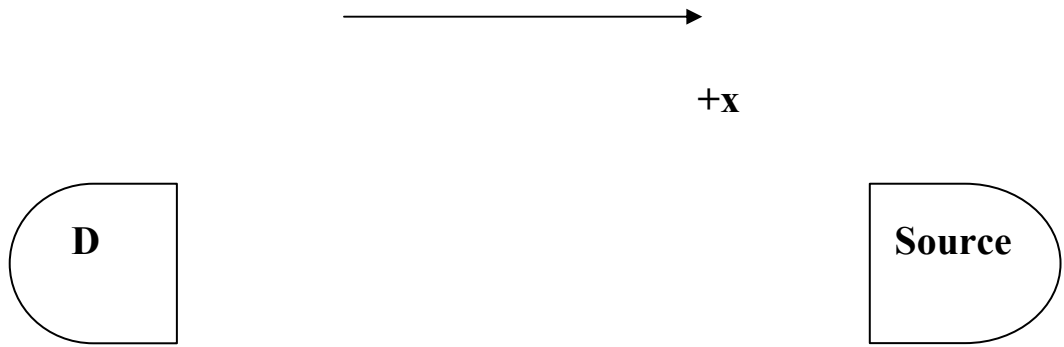

**Figure 10**

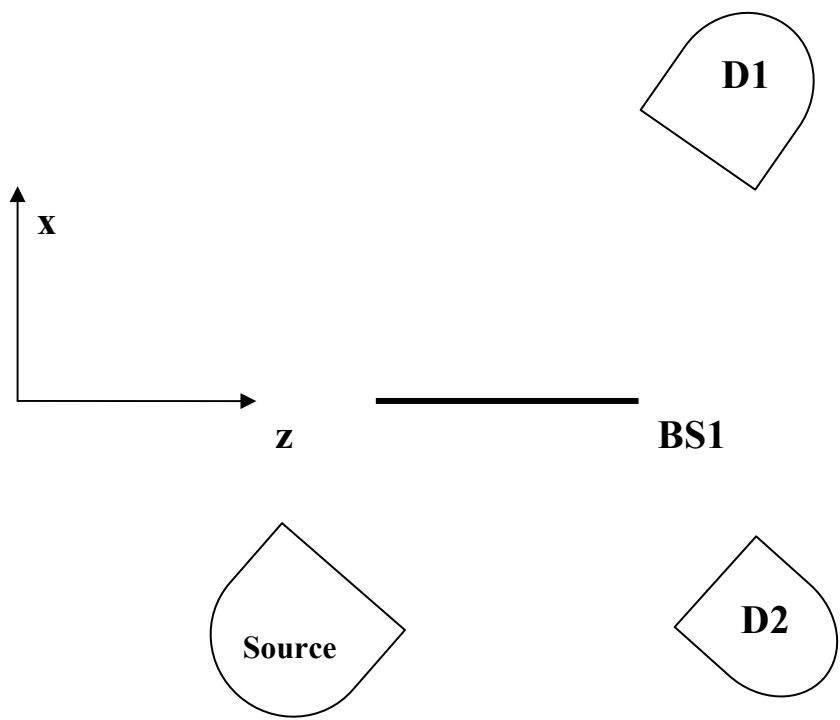

**Figure 11**

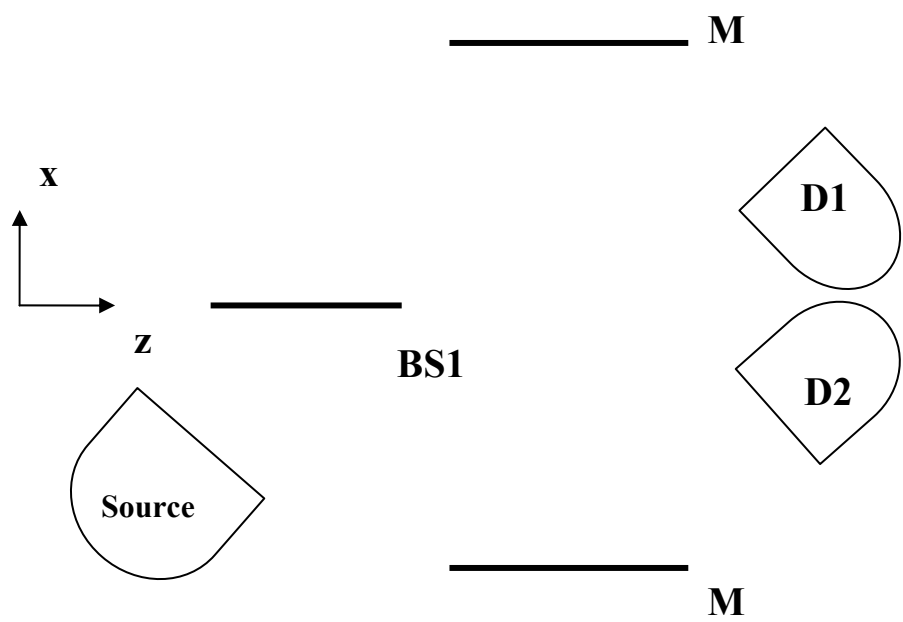

**Figure 12**

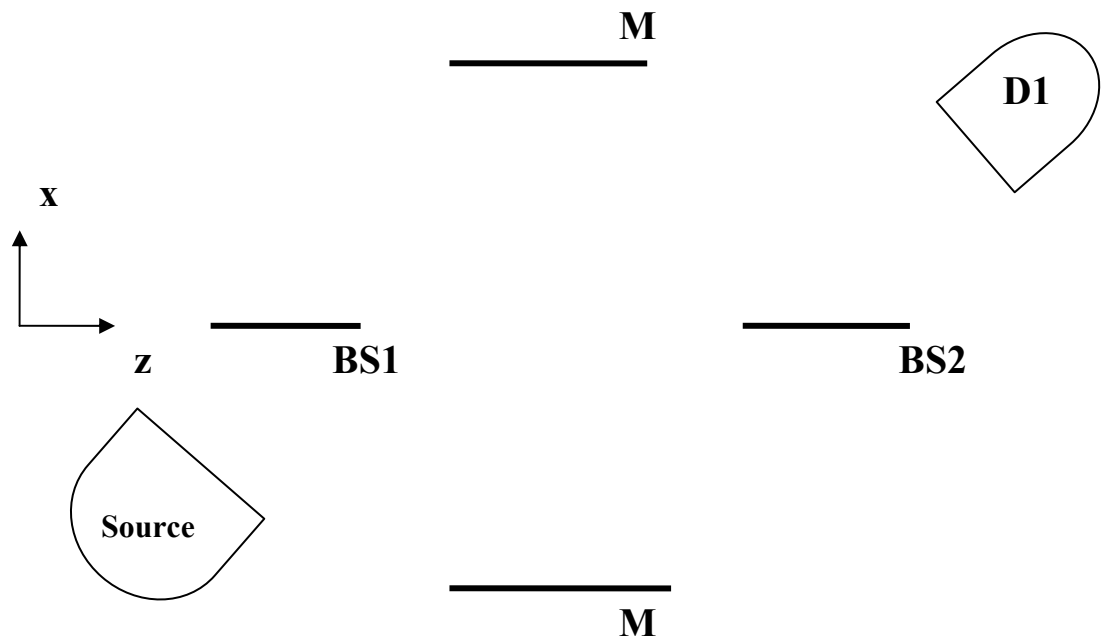

**Figure 13**

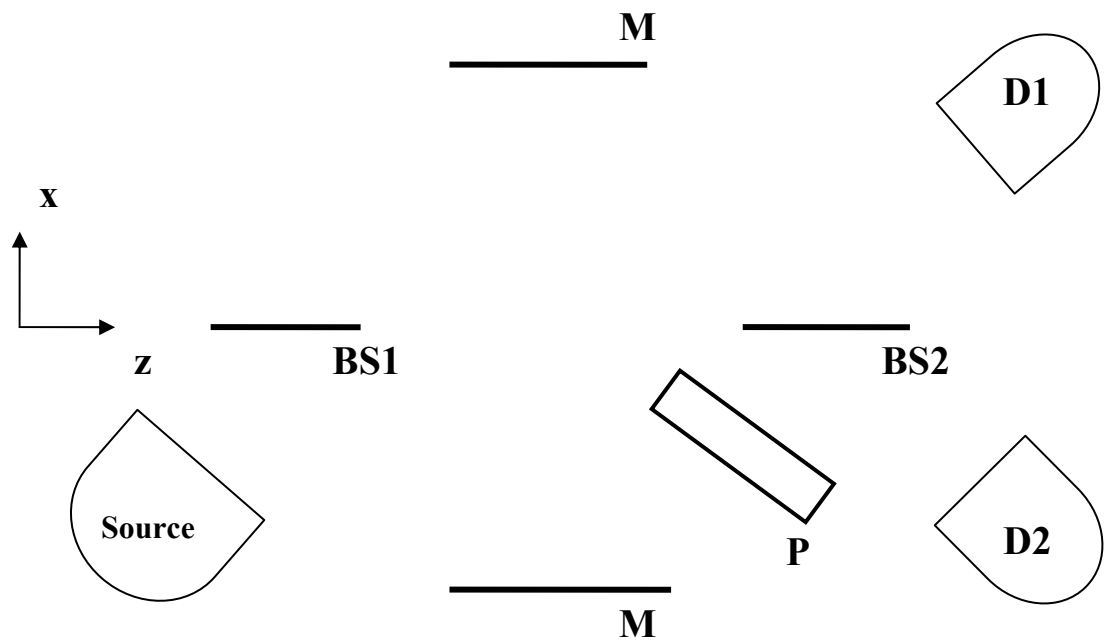

**NOTES & REFERENCES**

[13] This view differs from Mermin (1998) in that *spacetime relations* are fundamental in RBW while *quantum correlations* are fundamental in the Ithaca Interpretation.

[14] Kaiser (1981 & 1990).

[15] Kaiser (1981), p. 706.

[16] Bohr & Ulfbeck (1995), section D of part IV, p. 28.

[17] Ibid., p. 24.

[18] It is interesting to note that the coordinate transformations responsible for the canonical commutation relations (see Appendix B) differ from those of Galilean transformations by a time shift. This time shift is responsible for RoS which is "essential" to the kinematical perspective of RBW, but it introduces only a phase factor which is "without effect" in the dynamical perspective. See Bohr & Ulfbeck (1995), section D.6 of part IV, p. 29. See also J.V. Lepore, "Commutation Relations of Quantum Mechanics," *Phys. Rev.* **119**, 821-826 (1960) and H.R. Brown & P.R. Holland, "The Galilean covariance of quantum mechanics in the case of external fields," *Am. J. Phys*. **67**, 204-214 (1999).

[19] Bohr, Mottleson & Ulfbeck (2004).

[20] Personal correspondence to W. M. Stuckey dated 22 Dec 2004.

[21] Bohr & Ulfbeck (1995), p. 3.

[22] Bohr, Mottelson & Ulfbeck (2004), p. 411.

[23] Bohr & Ulfbeck (1995).

[24] Ibid

[25] Ibid., p. 27

[26] Howard Georgi, *Lie Algebras in Particle Physics*, 2nd Ed (Perseus Books, 1999), p. 14.

[27] Bohr, Mottelson & Ulfbeck (2004).

[28] Georgi (1999), p. 14.

[29] Ibid., p. 25.

[30] Ibid., p. 18.